\documentclass[final]{aipproc}

\layoutstyle{8x11double}

\usepackage{amssymb,amsmath,amsfonts,latexsym,graphicx,epsfig,here}

\newcommand{\beaa}{\begin{eqnarray*}} 
\newcommand{\enaa}{\end{eqnarray*}}

\newcommand{\bea}{\begin{eqnarray}}
\newcommand{\ena}{\end{eqnarray}} 

\newcommand{\be}{\begin{equation}}
\newcommand{\en}{\end{equation}}

\newcommand{\nn}{\nonumber\\}

\begin{document}

\title{A relativistic quark model with infrared confinement \\ and the 
tetraquark state}

\classification{12.39.Ki,13.25.Ft,13.25.Jx,14.40.Rt}

\keywords      {Relativistic quark model, confinement, exotic mesons}

\author{S.~Dubnicka}{
  address={Institute of Physics
Slovak Academy of Sciences
Dubravska cesta 9
SK-842 28 Bratislava, Slovak Republic}
}

\author{A.Z.~Dubnickova}{
  address={Comenius University
Dept. of Theoretical Physics
Mlynska Dolina
SK-84848 Bratislava, Slovak Republic}
}

\author{M.A.~Ivanov}{
  address={Bogoliubov Laboratory of Theoretical Physics, 
Joint Institute for Nuclear Research, 141980 Dubna, Russia}
}

\author{J.G.~K\"{o}rner}{
  address={Institut f\"{u}r Physik, Johannes Gutenberg-Universit\"{a}t,
D--55099 Mainz, Germany}
}

\author{ G.G.~Saidullaeva}{
  address={Al-Farabi Kazak National University, 480012 Almaty, Kazakhstan}
}

\begin{abstract}
We explore the consequences of treating the X(3872) meson as a tetraquark
bound state. As dynamical framework we employ a relativistic constituent 
quark model which includes infrared confinement in an effective way. 
We calculate the decay widths of the observed
channels $X\to J/\psi+2\pi (3\pi)$ and $X\to \bar D^0 +  D^0+\pi^0$ via the
intermediate off--shell states $X\to J/\psi+\rho(\omega)$ and 
$X\to \bar D +  D^\ast$.
For reasonable values of the size parameter $\Lambda_X$ of the 
X(3872) we find consistency with the available experimental data.

\end{abstract}

\maketitle

%%%%%%%%%%%%%%%%%%%%%%%%%%%%%%%%%%%%%%%%%%%%
%% MAINMATTER
%%%%%%%%%%%%%%%%%%%%%%%%%%%%%%%%%%%%%%%%%%%%

\section{Introduction}
\label{sec:intro}

A narrow charmonium--like state $X(3872)$ was observed in 2003 
in the exclusive decay process $B^\pm\to K^\pm\pi^+\pi^-J/\psi$
\cite{Choi:2003ue}. 
The  $X(3872)$ decays into $\pi^+\pi^-J/\psi$ and has a mass of 
$m_X=3872.0 \pm 0.6 ({\rm stat}) \pm 0.5 ({\rm syst}) $
very close to the $M_{D^0}+M_{D^{\ast\,0}}=3871.81 \pm 0.25$ mass threshold 
\cite{PDG}.
Its width was found to be less than 2.3 MeV at $90\%$ confidence level.
The state was confirmed in B-decays by the BaBar experiment 
\cite{Aubert:2004fc} 
and in $p\overline{p}$ production
by the Tevatron experiments \cite{Tevatron}. 

From the  observation of the decay $X(3872)\rightarrow J/\psi \gamma$ reported 
by \cite{jpsigamma}, 
it was shown that the only quantum numbers 
compatible with the data are $J^{PC}=1^{++}$ or $2^{-+}$. 
However, the observation of the decays into 
$D^0\overline{D}^{0}\pi^0$ by the Belle and BaBar collaborations 
\cite{jpsiDD} allows one to exclude the choice $2^{-+}$
because the near-threshold decay 
$X\to D^0\overline{D}^{0}\pi^0$   is expected to be strongly suppressed 
for $J=2$.

The Belle collaboration has  reported evidence for the decay mode  
$X \to \pi^+\pi^-\pi^0 J/\psi$ with  a strong three-pion peak between 750 MeV
and the kinematic limit of 775 MeV \cite{jpsigamma}, suggesting that 
the process is dominated by the  sub-threshold decay $X \to \omega J/\psi$. 
It was found that the branching ratio
of this mode is almost the same as that of the mode $X \to \pi^+\pi^- J/\psi$:
\be
\hspace*{-0.5cm}
\frac{ {\cal B}(X\to J/\psi\pi^+\pi^-\pi^0) }
     { {\cal B}(X\to J/\psi\pi^+\pi^-) }
 = 1.0 \pm 0.4 ({\rm stat}) \pm 0.3 ( {\rm syst} ).
\label{eq:expt}
\en
These observations imply strong isospin violation because the three-pion decay
proceeds via an intermediate $\omega$-meson with isospin 0 whereas
the two-pion decay proceeds via the intermediate $\rho$-meson with isospin 1.
Also the two-pion decay via the intermediate $\rho$-meson is very difficult
to explain by using an interpretation of the $X(3872)$ as a
simple $c\bar c$ charmonium state with isospin 0.  

There are several different interpretations of the $X(3872)$ in the literature:
a molecule bound state ($D^0\overline{D}^{\ast\,0}$)  with small binding energy, 
a tetraquark state composed of a diquark and antidiquark,
threshold cusps,  hybrids  and glueballs.
A description of the current theoretical and experimental situation for the 
new charmonium states may be found in the reviews \cite{review-X}. 

We provided in Ref.~\cite{Dubnicka:2010kz}
an independent analysis of the the properties of the 
X(3872) meson which we interpret as a tetraquark state as in 
\cite{Maiani:2004vq}.
We worked in the framework of the relativistic constituent quark model which 
has recently been extended to include infrared confinement effects 
\cite{Branz:2009cd}. 

\section{Theoretical framework}

The authors of \cite{Maiani:2004vq} suggested to consider the $X(3872)$ meson
as a $J^{PC}=1^{++}$ tetraquark state with a symmetric spin distribution:
$[cq]_{S=0}\,[\bar c \bar q]_{S=1} + [cq]_{S=1}\,[\bar c \bar q]_{S=0}$,
$(q=u,d)$. The nonlocal version of the four-quark interpolating current
reads
\bea
&&
J^\mu_{X_q}(x) = \int\! dx_1\ldots \int\! dx_4 
\delta\left(x-\sum\limits_{i=1}^4 w_i x_i\right) 
\label{eq:cur}\\
&&
\times\Phi_X\Big(\sum\limits_{i<j} (x_i-x_j)^2 \Big)
\nn
%%%%%%%%%%%%%
&&
\times
\frac{1}{\sqrt{2}}\, \varepsilon_{abc}\varepsilon_{dec} \,
\Big\{\, [q_a(x_4)C\gamma^5 c_b(x_1)][\bar q_d(x_3)\gamma^\mu C \bar c_e(x_2)]
\Big.
\nn
&&
\Big.
 + (\gamma^5 \leftrightarrow \gamma^\mu)
%      +[q_a(x_4)C\gamma^\mu c_b(x_1)][\bar q_d(x_3)\gamma^5 C \bar c_e(x_2)]
\,\Big\},
\nonumber
\ena 
where $w_1 = w_2 = m_c/2(m_q+m_c)$
and
$w_3 = w_4 = m_q/2(m_q+m_c)$.
The matrix $C=\gamma^0\gamma^2$ is the charge conjugation matrix.
The effective interaction Lagrangian describing the coupling
of the meson $X_q$ to its constituent quarks is written in the form
\be
{\cal L}_{\rm int} = g_X\,X_{q\,\mu}(x)\cdot J^\mu_{X_q}(x), \qquad (q=u,d).
\label{eq:lag}
\en     
The state $X_u$ breaks isospin symmetry maximally so
the authors of \cite{Maiani:2004vq} take the physical states to be a linear
superposition of the $X_u$ and $X_d$ states according to
\bea  
X_l\equiv X_{\rm low} &=&\hspace{0.2cm}  X_u\, \cos\theta +  X_d\, \sin\theta,\nn
X_h\equiv X_{\rm high} &=& - X_u\, \sin\theta +  X_d\, \cos\theta.
\label{eq:mixing}
\ena
The mixing angle $\theta$ can be determined from fitting the ratio
of branching ratios Eq.~(\ref{eq:expt}).

\begin{figure}[htbp]
\includegraphics[scale=0.3]{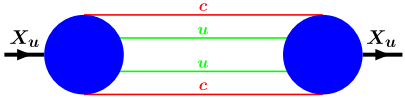} 
\caption{Diagram describing the $X_{u}$-meson mass operator.}
\label{fig:mass}
\end{figure}

The coupling constant $g_X$ in Eq.~(\ref{eq:lag}) will be determined from
the compositeness condition $Z_{H}=0$, see, e.g. Refs.~\cite{Efimov:1993zg}
and \cite{SWH}.
It gives
\be
\label{eq:Z=0}
Z_X = 1-\Pi_X^\prime(m^2_X)=0,
\en
where $\Pi_X(p^2)$ is the scalar part of the vector-meson mass operator.
The corresponding three-loop diagram describing the X-meson mass operator is 
shown in  Fig.~\ref{fig:mass}.
We will choose a simple Gaussian form for the vertex function
with the only dimensional parameter $\Lambda_X$ characterizing the size 
of the X-meson.

In \cite{Branz:2009cd} we described how to integrate $n$-point one-loop 
diagrams and how to implement infrared confinement of quarks in this process. 
We extend our loop integration techniques to the case of arbitrary
number of loops. Let $n$, $\ell$ and  $m$ be the number
of the propagators, loops and vertices, respectively.
In Minkowski space the $\ell$-loop diagram will be represented as
\bea
&&
\Pi(p_1,...,p_m) = 
\nn
&=&
\int\!\! [d^4k]^\ell  
\prod\limits_{i_1=1}^{m} \,
\Phi_{i_1+n} \left( -K^2_{i_1+n}\right)
\prod\limits_{i_3=1}^n\, S_{i_3}(\tilde k_{i_3}+v_{i_3}),
\nn
&&\nn
&&
K^2_{i_1+n} =\sum_{i_2}(\tilde k^{(i_2)}_{i_1+n}+v^{(i_2)}_{i_1+n})^2
\label{eq:diag}
\ena
where the vectors $\tilde k_i$  are  linear combinations 
of the loop momenta $k_i$. The $v_i$ are  linear combinations 
of the external momenta $p_i$ to be specified in the following.
The strings of Dirac matrices appearing in the calculation need not concern 
us since they do not depend on the momenta. 
The external momenta $p_i$ are all chosen to be ingoing such that one has 
$\sum\limits_{i=1}^m p_i=0$. 
 
Using the Schwinger representation of the local quark propagator one has
\be
S(k) = (m+\not\! k)
\int\limits_0^\infty\! 
d\beta\,e^{-\beta\,(m^2-k^2)}\,.
\en
The integrand in Eq.~(\ref{eq:diag})
has a Gaussian form  and may be integrated out explicitly.
 After doing the loop integrations one obtains
\be
\Pi =  \int\limits_0^\infty d^n \beta \, F(\beta_1,\ldots,\beta_n) \,,
\en
where $F$ stands for the whole structure of a given diagram. 
The set of Schwinger parameters $\beta_i$ can be turned into a simplex by 
introducing an additional $t$--integration  leading to 
\be
\hspace*{-0.2cm}
\Pi   = \int\limits_0^\infty\! dt t^{n-1}\!\! \int\limits_0^1\! d^n \alpha \, 
\delta\Big(1 - \sum\limits_{i=1}^n \alpha_i \Big) \, 
F(t\alpha_1,\ldots,t\alpha_n). 
\label{eq:loop_2} 
\en
There are altogether $n$ numerical integrations: $(n-1)$ $\alpha$--parameter
integrations and the integration over the scale parameter $t$. 
The very large $t$-region corresponds to the region where the singularities
of the diagram with its local quark propagators start appearing. 
However, as described in \cite{Branz:2009cd}, if one introduces 
an infrared cut-off on the upper limit of the t-integration, all 
singularities vanish because the integral is now convergent for any value
of the set of kinematic variables.
By introducing the infrared cut-off one has removed all potential thresholds 
in the quark loop diagram, i.e. the quarks are never on-shell and are thus
effectively confined. We take the cut-off parameter $\lambda$ to be the 
same in all physical processes.

\begin{figure}[htbp]
\begin{tabular}{lr}
\includegraphics[scale=0.2]{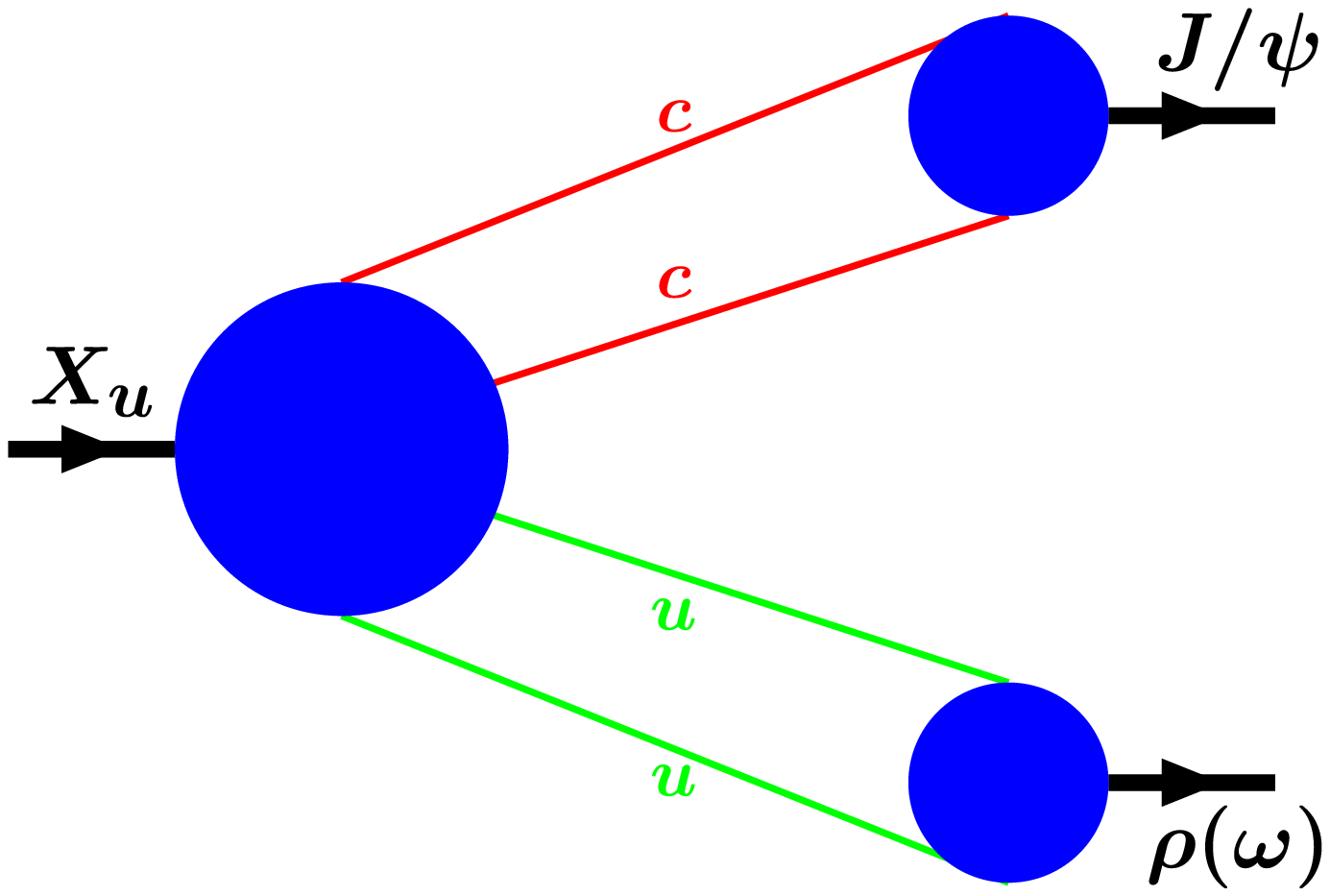} &
\includegraphics[scale=0.2]{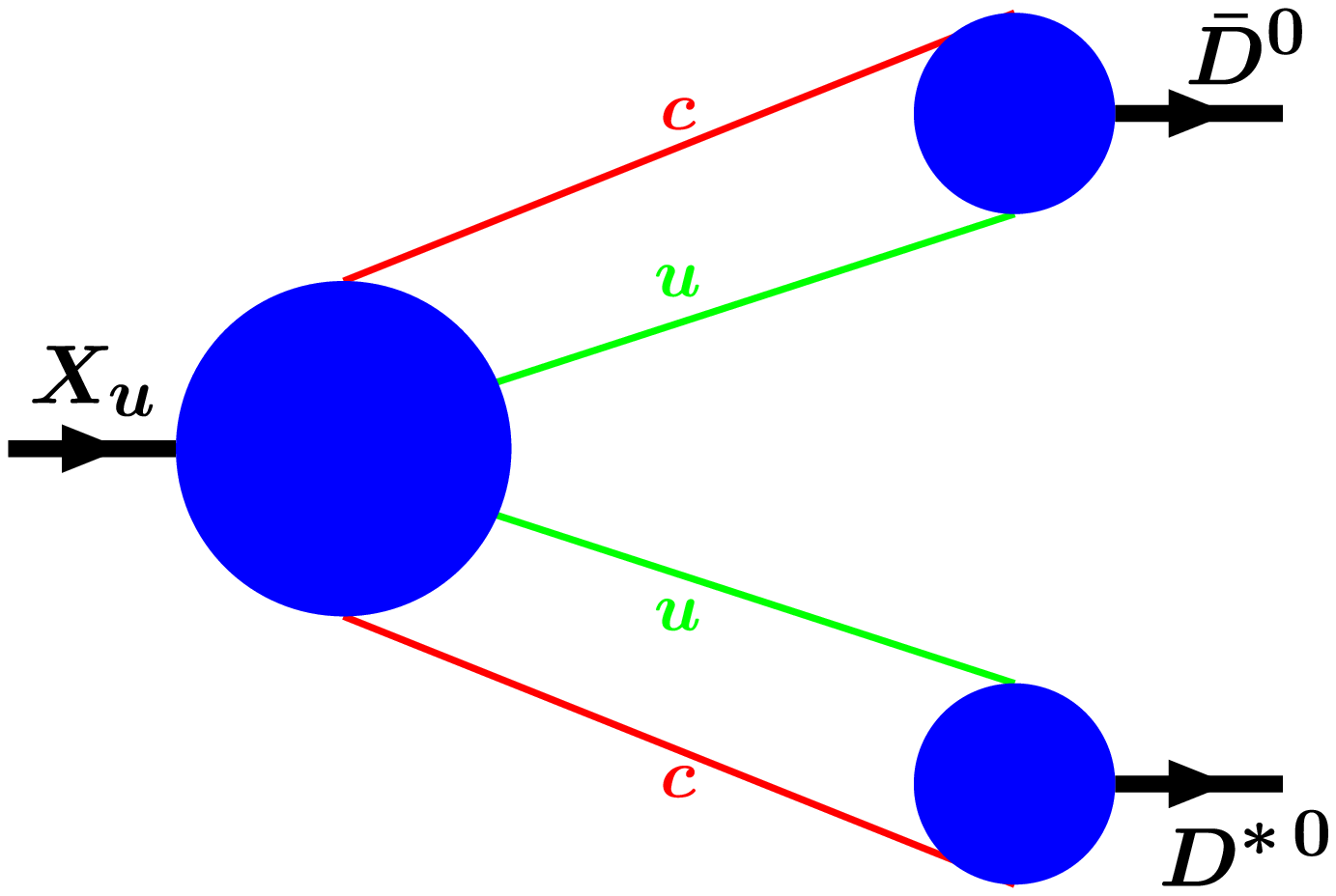}
\end{tabular}
\caption{
Feynman diagrams describing the decays
$X\to J/\psi+\rho(\omega)$ and $X\to D+\bar D^\ast$.
}
\label{fig:decay}
\end{figure}

Next we evaluate the matrix elements of the transitions
$X\to J/\psi+\rho(\omega)$ and $X\to D+\bar D^\ast$. 
The relevant Feynman diagrams are shown in Fig.~\ref{fig:decay}.
Since the X(3872) is very close to the respective thresholds in both cases, 
the intermediate $\rho$, $\omega$ and $D^\ast$ mesons 
have to be treated as off-shell particles.

\section{Numerical analysis}

Using the calculated matrix elements for the decay 
$X\to J/\psi+\rho(\omega)$ one can evaluate the decay widths
$X\to J/\psi+2\pi(3\pi) $. We employ the narrow width approximation
for this purpose.

The adjustable parameters of our model are the constituent quark masses $m_q$,
the scale parameter $\lambda$ characterizing the infrared confinement
and the size parameters $\Lambda_M$. They were determined  by using
a least square fit to a number of physical observables, see \cite{Branz:2009cd}.

There are two new free parameters: the mixing angle
$\theta$ in Eq.~(\ref{eq:mixing}) and the size parameter 
$\Lambda_X$. We have varied the parameter $\Lambda_X$
in a large interval and found that the ratio
\be
\frac{\Gamma(X_u\to J/\psi+3\,\pi)} 
     {\Gamma(X_u\to J/\psi+2\,\pi)} \approx 0.25
\en  
is very stable under variations of $\Lambda_X$. Hence, by using this result and
the central value of the experimental data given in Eq.~(\ref{eq:expt}),
one finds $\theta\approx \pm 18.4^{\rm o}$ for $X_l$~("+") and  $X_h$~("-"),
respectively. This is in agreement with the results obtained in both
\cite{Maiani:2004vq}: $\theta\approx \pm 20^{\rm o}$ and
\cite{Navarra:2006nd}: $\theta\approx \pm 23.5^{\rm o}$.
The decay width is quite sensitive to the change of the size
parameter $\Lambda_X$. A natural choice is to take a value 
close to $\Lambda_{J/\psi}$ and  $\Lambda_{\eta_c}$ 
which are both around 3 GeV.
We have varied the size parameter $\Lambda_X$ from  3 up to 4 GeV 
and found that
the decay width $\Gamma(X\to J/\psi+n\,\pi)$ decreases  from  0.30 
up to 0.07 MeV, monotonously. This result is in accordance with 
the experimental bound  $\Gamma(X(3872))\le 2.3$~MeV and the result
obtained in \cite{Maiani:2004vq}: 1.6 MeV.

In a similar way we calculate the width of the decay 
$X\to D^0\bar D^0\pi^0 $ which was observed by the Belle Coll.
and reported in \cite{jpsiDD}. 
We have  varied  $\Lambda_X$ from  3 up to 4 GeV and found that
the decay width $\Gamma(X_l\to \bar D^0 D^0 \pi^0)$ decreases  from  1.88 
up to 0.41 MeV, monotonously. 

Using the results of \cite{PDG}, one calculates the experimental rate ratio
\be
\frac{\Gamma(X\to D^0\bar D^0 \pi^0)}
     {\Gamma(X\to J/\psi\pi^+\pi^-)}  = 10.5\pm 4.7
\label{eq:ratio}
\en
The theoretical value for this rate ratio depends only weakly on the size 
parameter $\Lambda_{X}$

\be
\frac{\Gamma(X\to D^0\bar D^0 \pi^0)}
     {\Gamma(X\to J/\psi\pi^+\pi^-)}\Big|_{\rm theor}  = 6.0 \pm 0.2.
\en
The theoretical error reflects the $\Lambda_{X}$ dependence of the 
ratio. The ratio lies within the experimental
uncertainties given by Eq.~(\ref{eq:ratio}).

%%%%%%%%%%%%%%%%%%%%%%%%%%%%%%%%%%%%%%%%%%%%%%%%
%% BACKMATTER
%%%%%%%%%%%%%%%%%%%%%%%%%%%%%%%%%%%%%%%%%%%%%%%%

\begin{theacknowledgments}
This work was supported by 
the DFG grant KO 1069/13-1,
the Heisenberg-Landau program,  
the Slovak aimed project at JINR and the grant VEGA No.2/0009/10.
M.A.I. also appreciates the partial support of the 
Russian Fund of Basic Research grant No. 10-02-00368-a. 
\end{theacknowledgments}

\end{document}